\documentstyle[preprint,aps]{revtex}
\input epsf.tex
\def\DESepsf(#1 width #2){\epsfxsize=#2 \epsfbox{#1}}
\begin{document}
\preprint{\vbox{\hbox{}}}
\draft
\title{
Semi-Inclusive $B\to K(K^*) X$ Decays with Initial Bound State Effects}
\author{Xiao-Gang He$^1$, Changhao Jin$^2$ and J.P. Ma$^3$}
\address{
$^1$Department of Physics, National Taiwan University,
Taipei\\
$^2$School of Physics, University of Melbourne, Melbourne\\
and\\
$^3$Institute of Theoretical Physics, Academia Sinica, Beijing\\
}
\date{ November, 2000}
\maketitle
\begin{abstract}
The effects of initial $b$ quark bound state for the semi-inclusive decays
$B\to K(K^*) X$ are studied using light cone expansion
and heavy quark effective theory methods.
We find that the
initial bound state effects on the branching ratios and
CP asymmetries are small.
In the light cone expansion approach, the CP-averaged branching ratios are
increased by about 2\% with respect to the free $b$-quark decay.
For $\bar B^0 \to K^- (K^{*-}) X$, the CP-averaged branching ratios
are sensitive to the phase $\gamma$
and the CP asymmetry can be as large as $7\%$ ($14\%$),
whereas for $B^-\to \bar K^0 (\bar K^{*0})X$ the CP-averaged branching ratios
are not sensitive to $\gamma$ and the CP asymmetries are small ($< 1\%$).
The CP-averaged branching ratios
are predicted to be in the ranges $(0.53 \sim 1.5)\times 10^{-4}$
[$(0.25 \sim 2.0)\times 10^{-4}$] for $\bar B^0 \to K^- (K^{*-})X$
and $(0.77 \sim 0.84)\times 10^{-4}$ [$(0.67 \sim 0.74)\times 10^{-4}$] for
$B^-\to \bar K^0 (\bar K^{*0}) X$, depending on the value of the CP violating
phase $\gamma$. In the heavy quark effective theory approach,
we find that the branching ratios are decreased by about 10\% and the CP
asymmetries are not affected.
These predictions can be tested in the near future.
\end{abstract}


\preprint{\vbox{\hbox{}}}

\section{Introduction}

There have been considerable experimental and theoretical efforts to
understand the properties of B decays. These studies have provided important
information about
the mechanism for B decays and the origin of CP violation.
In the next few years large quantities of
experimental data on B decays will become available.
It is hoped that one will obtain
even more important
information in understanding
the mechanism for B decays and the mechanism for CP violation.
In particular, charmless hadronic B decays have played an important role in
the determination of
the CP violating parameter $\gamma$ in the Standard Model (SM)\cite{1,2,3,4,5}.
While most of the studies have concentrated on the exclusive B decay modes for
CP violation, there are also some studies for semi-inclusive decays\cite{2,3}.
At the quark level the relevant Hamiltonian for B decays in the SM is
well understood. The major uncertainties for these decays come from our
insufficient understanding of the long distance strong interaction dynamics
involved in these decays. There are several methods which have been used to
estimate the decay amplitudes,
including naive factorization, QCD improved factorization and
methods based on symmetry considerations.

Recently it has been argued that in the heavy quark limit, factorization
is a good approximation\cite{4} and several processes have been
calculated\cite{5}.
Leading QCD corrections to the naive factorization
can be studied for exclusive decays in a systematic way.
In the calculation of exclusive decays,
the hadronic matrix elements can be factorized and strong interaction dynamics can be
parameterized into the relevant decay constants, light cone distribution amplitudes
 and transition form factors.
At the present time, the light cone distribution amplitudes and transition form factors are
not well known which introduce uncertainties
in the calculations. Of course one should keep in mind that there may be large
corrections of order $\Lambda_{QCD}/m_b$ 
which needs further study.  From quark hadron 
duality consideration, inclusive decays can
be represented by quark level calculations and the uncertainties may be small.
It is believed that theoretical calculations
for exclusive decays contain more uncertainties than inclusive decays. Of course
when going completely inclusive,
there are less information that can be extracted
about strong and
weak interaction dynamics and CP violation, and it is experimentally hard to identify
final states inclusively.
In this paper we will take the
way in between by studying semi-inclusive decays following Ref.\cite{2} in the hope that
one may be able to reduce some of the hadronic uncertainties in exclusive decays
on one hand, and
still be able to obtain important information about B decays and CP violation with
clear experimental signal on the other.
We will study the charmless semi-inclusive decays $B\to K X$ and $B \to K^* X$.
Here the $X$ indicates states containing no charmed particles.

The decay modes $B\to K(K^*) X$ have been studied before\cite{2,3}.
In previous studies,
several effects were treated phenomenologically, such as the number of colors
was taken as an effective number and treated as a free parameter, the gluon
virtuality $q^2$ in the penguin diagrams was assumed to be around $m^2_b/2$, and
the bound state effects of b-quark inside the $B$ meson was modeled by assuming
its momentum to obey a Gaussian distribution.
To have a better understanding of these decays,
it is necessary to carry out calculations in such a way that the
phenomenological treatments can be improved with better theoretical
understanding.
It has recently been shown that it is indeed possible in the heavy quark limit
to handle most of the problems in exclusive $B$ to two light meson decays from
QCD calculations\cite{4}.
We will use the same formalism in
our study of semi-inclusive decays in the factorization approximation, paying
particular attention to the initial bound state effects.

The problems treated in the case of exclusive decays are
different in some ways from the semi-inclusive
decays studied here. The problems associated with the number of colors and
the gluon virtuality can be treated the same way, but the initial b quark bound 
effects in semi-inclusive decays arise in different form from those in exclusive decays.
In the exclusive decay case, the $b$ quark bound state effects are taken
care by decay constants and transition form factors. In the semi-inclusive
case, there are contributions which, in the free quark decay approximation,
can be viewed as a $b$ quark decay into a meson and another quark. One needs to
treat initial $b$ quark bound state effects on more theoretical ground.
This will be the main focus of this paper. We will study this problem
using two different methods with one based on light cone expansion and
another based on heavy quark effective theory.

To further reduce possible uncertainties associated with form factors,
we will choose processes which have the
least numbers of hadronic parameters beside
the ones related to the initial bound state effects.
We find that the following
processes are particularly good for this purpose,

\begin{eqnarray}
&&\bar B^0 \to K^- X,\;\;\;\;B^- \to \bar K^0 X,\nonumber\\
&&\bar B^0 \to K^{*-} X,\;\;\;\;B^-\to \bar K^{*0} X.
\end{eqnarray}
For these processes, the transition form factors for $B\to K$ and
$B\to K^*$ do not show up
in the factorization approximation because the bi-quark
operator $\bar s \Gamma b$ (here $\Gamma$ is some appropriate Dirac
matrices) does not change the electric charge of the
initial particle $B$ and the final particle $K(K^*)$.
Therefore for these processes there are only the $K(K^*)$ decay constants
and parameters related to
the initial bound state effects if small annihilation contributions are
neglected.

The paper is arranged as follows.
In Section II, we will study the decay amplitudes
in the SM for the semi-inclusive $B\to K(K^*) X$ decays. In
Section III, we will study the light cone and heavy quark effective theory
formulation of
the initial bound state effects on these semi-inclusive decays. And in
Section IV, we will carry out numerical analyses of the energy spectra of the
$K(K^*)$, branching ratios and
CP asymmetries in $B\to K(K^*) X$, and draw our conclusions.

\section{Decay amplitudes in the heavy quark limit}

In this section we study the short distance decay amplitudes for
semi-inclusive $B\to K(K^*) X$ decays.
The effective Hamiltonian for charmless B decays with
$\Delta S = 1$ at the quark level is given by

\begin{eqnarray}
H_{eff} &=&{G_F\over \sqrt{2}}
\left \{ V_{ub} V^*_{us}(c_1 O_1 +c_2 O_2 + \sum_{n=3}^{11} c_n O_n)
+  V_{cb} V^*_{cs} \sum_{n=3}^{11} c_n O_n\right \}.
\label{hamiltonian}
\end{eqnarray}
Here $O_n$ are quark and gluon operators and are given by

\begin{eqnarray}
&&O_1 = (\bar s_i u_j)_{V-A} (\bar u_j b_i)_{V-A},\;\;
O_2 = (\bar s_i u_i)_{V-A} (\bar u_j b_j)_{V-A},\nonumber\\
&&O_{3(5)} = (\bar s_i b_i)_{V-A}\sum_{q'}
(\bar q^\prime_j q^\prime_j)_{V-(+)A},\;\;
O_{4(6)} = (\bar s_i b_j)_{V-A}\sum_{q'}
(\bar q^\prime_j q^\prime_i)_{V-(+)A},\nonumber\\
&&O_{7(9)} = {3\over 2}(\bar s_i b_i)_{V-A}\sum_{q'}
e_{q^\prime}(\bar q^\prime_j q^\prime_j)_{V+(-)A},\;\;
O_{8(10)} ={3\over 2} (\bar s_i b_j)_{V-A}\sum_{q'}
e_{q^\prime}(\bar q^\prime_j q^\prime_i)_{V+(-)A},\nonumber\\
&&O_{11} = {g_s\over 8\pi^2} m_b \bar s_i \sigma^{\mu\nu} G_{\mu\nu}^a
{\lambda_a^{ij}\over 2}(1+\gamma_5)b_j,
\end{eqnarray}
where $(V\pm A)(V\pm A) =\gamma^\mu(1\pm\gamma_5) \gamma_\mu(1\pm \gamma_5)$,
$q^\prime = u,d,s,c,b$, $e_{q^\prime}$ is the electric charge number
of the $q^\prime$ quark, $\lambda_a$ is the color SU(3) Gell-Mann matrix,
$i$ and $j$ are color indices, and
$G_{\mu\nu}$ is the gluon field strength.

The Wilson coefficients $c_n$ have been calculated in
different schemes\cite{6}. In this paper we will use consistently the NDR
scheme. The values of $c_n$ at $\mu \approx m_b$ with the
next-to-leading order (NLO) QCD corrections are given by\cite{6}

\begin{eqnarray}
&&c_1 = -0.185,\;\;c_2 = 1.082,\;\;c_3=0.014,\;\;c_4 = -0.035,\;\;
c_5=0.009,\;\;c_6 = -0.041,\nonumber\\
&&c_7= -0.002\alpha_{em},\;\;c_8=0.054\alpha_{em},\;\;
c_9=-1.292\alpha_{em},\;\;c_{10}=-0.263\alpha_{em},\;\;
c_{11} = -0.143.\nonumber
\end{eqnarray}
Here $\alpha_{em}=1/137 $ is the electromagnetic fine structure constant.

In order to make sure that the observed events are from rare charmless B decays,
and other processes, such as $B\to D(D^*) X' \to K (K^*) X''$,
do not contaminate
the direct rare decay of $B\to K(K^*) X$ due to short distance interaction,
we will make a cut on the $K(K^*)$ energy which will be set
at $E_{K,K^*} > 2.1$ GeV. It has been shown that this cut can eliminate
most of the unwanted events while leave most of the events induced by short
distance contributions\cite{2} because the matrix elements of the type $<K(K^*)|
j_1|0><X|j_2|B>$ would results in a fast $K(K^*)$ in the final state.
The resulting events will resemble two body type of decays
with one of them be the  $K(K^*)$ and another, back-to-back against
the $K(K^*)$,
will be $X$ with small invariant mass $M_X^2$. With the cut $E_{K,K^*}> 2.1$ GeV,
$M^2_X < 5.7$ GeV$^2$.

The hadronic matrix element for a specific operator $<X K|O|B>$ is difficult
to calculate at present. We will use factorization approximation to estimate
it. The factorization approximation has been shown to hold in the heavy quark
limit for exclusive $B$ decays into two light hadrons.
The leading contribution for an operator which can be written as
a product of two currents $j_1 = \bar s \Gamma_1 q'$ and $j_2 = \bar q'
\Gamma_2 b$ with $\Gamma_i$ carrying appropriate Lorentz and Dirac indices,
$O= j_1\cdot j_2$, is given by

\begin{eqnarray}
<X K| O|B>_{fact} &=& <K|j_1|0><X|j_2|B> + <X|j'_1|0><K|j'_2|B>\nonumber\\
& +& <X K|j_1|0><0|j_2|B>.
\label{facto}
\end{eqnarray}
The second term on the right-hand-side in the above represents the
Fierz transformed factorization
terms with $j'_1 = \bar q' \Gamma'_1 q'$ and $j'_2=\bar s \Gamma'_2 b$.
The third term is usually referred to as the annihilation contribution.

$B\to KX$ is a many-body decay, which is different from two-body decays.
There are more ways of factorization for a many-body decay,
such as
$<X_1 K|j_1|0><X_1'|j_2|B>$ and $<X_2|j'_1|0><X_2' K|j'_2|B>$,
with $X = X_1 +X_1' = X_2 +X_2'$.
The three terms in Eq.~(\ref{facto}) corresponding to
the cases: $<X_1| = <0|$, $<X_2'| = <0|$ and $<X_1'|=<0|$, respectively.
For $B\to KX$ with a cut $E_K > 2.1$ GeV, the final state $X$ has a small invariant mass.
This is a quasi-two-body
decay, with $K$ and $X$ moving rapidly
apart in opposite directions. The probability of forming the final state
$<X_1 K|$ with $<X_1|\neq <0|$ is less than the probability of forming the simple final state
$<K|$. This suggests that the contribution of the configuration
$<X_1 K|j_1|0><X'_1|j_2|B>$ is dominated by $<K|j_1|0><X|j_2|B>$. Likewise,
the contribution of the configuration $<X_2|j'_1|0><X'_2K|j'_2|B>$ is
dominated by $<X|j'_1|0><K|j'_2|B>$.
The cases with $|X_1>$ and $|X_2'>$ not equal to
$|0>$ are also higher order in $\alpha_s$ and therefore $\alpha_s$ power
suppressed. We will
neglect them in our later discussions which also eliminate the third term
in Eq.~(\ref{facto}).

The above approximation is also supported by explicit calculation of the bremsstrahlung process,
$b\to K q' g$, which represents some of the $\alpha_s$ order corrections.
It has been shown, in a similar situation of $b\to \phi s$ and $b\to \phi s g$,
that the
bremsstrahlung contributes less than 3\% of the total branching ratio\cite{7}.
One can easily obtain from
Ref.\cite{7} an estimate of the contribution for the processes considered here.
The bremsstrahlung contribution is small.
Eq.~(\ref{facto}) will adequately approximate the leading contributions and we
will work with this approximation.

In the heavy quark limit, a class of radiative corrections in powers of $\alpha_s$,
which does not change the form of the operators, can
be included for the  matrix elements. For a local
operator the correction can be parameterized as the following, similar to
the exclusive decays discussed in Refs. \cite{4,5},

\begin{eqnarray}
<X K|O|B> = <X K|O|B>_{fact} [1+ \sum_{n=1}^{\infty} r_n \alpha_s^n +
O(\Lambda_{QCD}/m_b)],
\label{fact}
\end{eqnarray}
where $<X K|O|B>_{fact}$ denotes the naive factorization result.
$\Lambda_{QCD}\approx 0.3$ GeV is the strong interaction scale.
The second and third terms in the square bracket indicate, respectively,
higher order $\alpha_s$ and $\Lambda_{QCD}/m_b$
corrections to the factorized matrix element.

Similar arguments can be made for $B\to K^*X$ decays also.
For the $<K(K^*)|j_1|0><X|j_2|B>$ type, the decay amplitudes
involves the $K(K^*)$ decay constants, while for the
$<X|j'_1|0><K(K^*)|j'_2|B>$ type, it
involves the transition form
factors from $B$ to $K(K^*)$, and the $<XK(K^*)|j_1|0><0|j_2|B>$ type
involves the $B$ decay constant.

If all three
terms in Eq.~(\ref{facto})
contribute with the same order of magnitude, the accumulated uncertainties
will be substantial due to large
uncertainties in the transition form
factors and the $B$ decay constant. Fortunately
we find that for $\bar B^0 \to K^- (K^{*-}) X$ and
$B^- \to \bar K^0 (\bar K^{*0}) X$, only
the first and the third types of terms in Eq.~(\ref{facto})
contribute due to electric charge conservation.
This eliminates possible uncertainties from the transition form factors.
Also as argued before
the third term can be neglected because it 
is subleading and $\alpha_s$ power suppressed.
There is only one term present, which considerably
simplifies the calculation.

Using the effective Hamiltonian in Eq.~(\ref{hamiltonian}),
we obtain

\begin{eqnarray}
A( B \to K X )&=&
i{G_F\over \sqrt{2}}\sum_{q=u,c} V_{qb}V_{qs}^*
f_K [A^q P_K^\mu <X|\bar q' \gamma_\mu (1-\gamma_5) b |B> \nonumber\\
&+& B^q <X|\bar q'(1-\gamma_5)b|B> ], \nonumber \\
A( B \to K^* X )&=&
{G_F\over \sqrt{2}}\sum_{q=u,c} V_{qb}V_{qs}^*
m_{K^*} f_{K^*}
\tilde A^q \epsilon_{\lambda}^{\mu*} <X|\bar q' \gamma_\mu (1-\gamma_5) b |B>,
\label{app}
\end{eqnarray}
where $q' = u$ and $d$ for $\bar B^0$ and $B^-$, respectively.
The decay constants are defined as
$<K|\bar s \gamma^\mu (1-\gamma_5)q'|0> = i f_K P_K^\mu$
and $<K^*(\lambda)|\bar s \gamma^\mu (1-\gamma_5)q'|0> =
m_{K^*} f_{K^*} \epsilon_{\lambda}^{\mu*}$.
We adopt the standard covariant normalization $<B|B> = 2E_B(2\pi)^3\delta^3({\bf 0})$.
The coefficients $A^q (\tilde A^q)$ and $B^q$
are given by, for $\bar B^0 \to K^-(K^{*-}) X$

\begin{eqnarray}
A^q (\tilde A^q) &=& a_1^q +a^q_4 + a_{10}^q + a_{10a}^q,\nonumber\\
B^q &=&
(a_6^q+a_8^q + a_{8a}^q)
{2m^2_{K^-}\over m_u+m_s}.
\end{eqnarray}
For $B^- \to \bar K^0 (\bar K^{*0}) X$,

\begin{eqnarray}
A^q(\tilde A^q) &=& a^q_4 -{1\over 2} a_{10}^q + a_{10a}^q,\nonumber\\
B^q &=&
(a_6^q-{1\over 2} a_8^q + a_{8a}^q)
{2m^2_{K^0}\over m_d+m_s}.
\end{eqnarray}

Including the lowest $\alpha_s$ order corrections in Eq.~(\ref{fact}), $a^q_i$
are given by

\begin{eqnarray}
&&a^u_1 = c_2 +{c_1\over N} + {\alpha_s \over 4\pi} {C_F\over N} c_1 F_P,\nonumber\\
&&a^c_1 = 0,\nonumber\\
&&a_4^q = c_4 + {c_3\over N}
+{\alpha_s\over 4\pi} {C_F\over N}
\left [ c_3(F_P + G_P(s_s) + G_P(s_b))
+ c_1 G_P(s_q)\right. \nonumber\\
 &&\;\;\;\;+ \left .
(c_4+c_6) \sum_{f=u}^b G_P(s_f) + c_{11}G_{P,11}\right ],
\nonumber\\
&&a_6^q = c_6 + {c_5\over N}
+{\alpha_s\over 4\pi} {C_F\over N}
\left [ c_3( G'_P(s_s) + G'_P(s_b))
+ c_1 G'_P(s_q)\right. \nonumber\\
 &&\;\;\;\;+ \left .
(c_4+c_6) \sum_{f=u}^b G'_P(s_f) + c_{11}G'_{P,11}\right ],
\nonumber\\
&&a_8^q = c_8 +{c_7\over N},\nonumber\\
&&a_{8a}^q=
{\alpha_s\over 4\pi} {C_F\over N}
\left [ (c_8+c_{10}) {3\over 2} \sum _{f=u}^be_f G'_P(s_f)
+c_9 {3\over 2} (e_s G'_P(s_s) + e_b G'_P(s_b))\right ],\nonumber\\
&&a_{10}^q = c_{10} +{c_{9}\over N} + {\alpha_s\over 4\pi}
{C_F\over N} c_{9}F_P,\nonumber\\
&&a_{10a}^q=
{\alpha_s\over 4\pi} {C_F\over N}
\left [ (c_8+c_{10}) {3\over 2} \sum _{f=u}^be_f G_P(s_f)
+c_9 {3\over 2} (e_s G_P(s_s) + e_b G_P(s_b))\right ],
\end{eqnarray}
where $N=3$ is the number of colors,
$C_F = (N^2-1)/(2N)$, and $s_f = m^2_f/m_b^2$.
The other items are given by

\begin{eqnarray}
&&F_P = -12 \ln{\mu \over m_b} - 18 + f^I_P,
\nonumber\\
&&f^I_P = \int^1_0 dx g(x) \phi_P(x),\;\;
g(x) =  3{1-2x\over 1-x} \ln x -3i\pi,\nonumber\\
&&G_P(s) = {2\over 3} - {4\over 3} \ln{\mu\over m_b}
+ 4\int^1_0 dx \phi_P(x) \int^1_0 du u(1-u) \ln[s-u(1-u)(1-x)-i \epsilon ],\nonumber\\
&&G_{P,11} = -\int^1_0 dx {2\over 1-x} \phi_P(x),\nonumber\\
&&G'_K(s) = {1\over 3} - \ln{\mu\over m_b}
+ 3\int^1_0 dx \phi^0_K(x) \int^1_0 du u(1-u) \ln[s-u(1-u)(1-x)- i\epsilon ],\nonumber\\
&&G'_{K,11} = -\int^1_0 dx {3\over 2} \phi^0_K(x),\nonumber\\
&&G'_{K^*}(s) = 0,\nonumber\\
&&G'_{K^*,11} = 0,
\end{eqnarray}
where the subscript $P$ can be $K$ or $K^*$, indicating
that the coefficients $a^q_i$ are process dependent.
$\phi_K(x)$ and $\phi^0_K(x)$ are the twist-2 and twist-3 kaon meson distribution
amplitudes, respectively. $\phi_{K^*}(x)$ is the leading twist distribution amplitude for
the longitudinally polarized $K^*$. In this paper we will
take the following forms for them\cite{5},

\begin{eqnarray}
&&\phi_{K,K^*}(x) = 6x(1-x),\;\;\;\;\phi^0_K(x) =1.
\end{eqnarray}

The amplitudes in Eq.~(\ref{app}) are from perturbative QCD calculation
in the heavy quark limit. The
number of colors should not be treated as an effective number, but has to be
3 from QCD. The results are, in principle, renormalization scale and scheme independent.
The problem associated with the gluon virtuality $k^2 =(1-x) m^2_B$
in the naive factorization calculation
is also meaningfully treated by convoluting the $x$-dependence with the
meson distribution amplitudes in the functions $G_P(s)$ and $G'_P(s)$.

\section{Initial bound state effects}

In this section we study the decay rates for $B\to K(K^*) X$, taking into
account $b$ quark bound state effects, using
two different methods, the light cone expansion method and the heavy quark
effective theory method.

We will work out, in detail, the formulation for $B\to K X$ in the following. The results
for $B\to K^* X$ can be easily obtained in a similar way.
Without taking into account the initial bound state effects,
that is in the free b quark decay approximation,
the decay can be viewed as the two body process
$b\to K q'$ and one obtains\cite{2}

\begin{eqnarray}
&&\Gamma(B\to K X)\approx \Gamma(b\to K q') = {f^2_K\over 8\pi}
(m_b^2 |\alpha|^2 + |\beta|^2 )m_b,\nonumber\\
&&\alpha = {G_F\over \sqrt{2}}\sum_{q=u,c} V_{qb}V_{qs}^*A^q,\;\;
\beta = {G_F\over \sqrt{2}} \sum_{q=u,c} V_{qb}V_{qs}^* B^q.
\label{free}
\end{eqnarray}

If the b quark mass is infinitively large, $Br(B\to   K(K^*) X)$ is
equal to $Br(b\to K(K^*) q')$. However due to initial b quark bound state
effects there are corrections\cite{8,9}.
We now proceed to study the initial bound state effects on the decay rates.

The differential decay rate for $B\to K X$ in the $B$ rest frame,
following the procedure in Ref.\cite{8}, is given by

\begin{eqnarray}
d\Gamma (B\to K X)
&=& {1\over 2m_B} {d^3 {\bf P}_K\over (2\pi)^3 2 E_K}
\sum_X (2\pi)^4 \delta^4(P_B - P_K - P_X) |A(B\to K X)|^2.
\label{rate}
\end{eqnarray}
Using $\int d^4y\,\, \mbox{exp}[-iy\cdot (P_B - P_K -P_X)] = (2\pi)^4 \delta^4(P_B-P_K-P_X)$, we have

\begin{eqnarray}
&&\sum_X (2\pi)^4 \delta^4(P_B-P_K-P_X)|A(B\to K X)|^2
=
f^2_K
\sum_X \int d^4y\,\, e^{-iy\cdot (P_B - P_K -P_X)}
\nonumber\\
&&\times [|\alpha|^2P^\mu_KP^\nu_K <B|\bar b \gamma_\nu (1-\gamma_5) q'
|X><X|\bar q' \gamma_\mu (1-\gamma_5) b|B>\nonumber\\
&&+|\beta|^2 <B|\bar b(1+\gamma_5) q'|X><X|\bar q'(1-\gamma_5) b|B>]\nonumber\\
&&=
f^2_K \int d^4y\,\, e^{iy\cdot P_K}
(|\alpha|^2 P_K^\mu P_K^\nu <B|[j^\dagger_\nu(0), j_\mu(y)]|B>
+|\beta|^2 <B|[J^\dagger(0), J(y)]|B>),
\end{eqnarray}
where $j_\mu = \bar q' \gamma_\mu (1-\gamma_5) b$ and $J = \bar q'(1-\gamma_5) b$.

Computing the current commutators one obtains

\begin{eqnarray}
&&\sum_X (2\pi)^4\delta^4(P_B-P_K-P_X)|A(B\to K X)|^2
\nonumber\\
&&= -2f_K^2 \left \{ |\alpha|^2 P_K^\mu P_K^\nu (g_{\mu \alpha}g_{\nu\beta}
+ g_{\mu\beta}g_{\nu\alpha} - g_{\mu\nu} g_{\alpha\beta})
+ |\beta|^2 g_{\alpha\beta}\right \}\nonumber\\
&&\times \int d^4y\,\, e^{iy\cdot P_K} [\partial^\alpha\Delta_{q'}(y)] <B|\bar b(0)
\gamma^\beta (1-\gamma_5) U(0,y) b(y)|B>.
\label{lc}
\end{eqnarray}
In the above we have assumed $m_{q'}=0$ and used

\begin{eqnarray}
\{q'(x), \bar q'(y)\} = i(\gamma\cdot \partial) i\Delta_{q'}(x-y) U(x,y),
\end{eqnarray}
with

\begin{eqnarray}
U(x,y) &=& {\cal P} \mbox{exp}[ig_s \int^x_y dz^\mu G_\mu(z)],\nonumber\\
\Delta_{q'}(y) &=& -{i\over (2\pi)^3} \int d^4k\,\, e^{-ik\cdot y}\epsilon(k^0)
\delta(k^2),
\end{eqnarray}
where $U(x,y)$ is the Wilson link, $G^\mu$ is the background gluon field,
and $\epsilon(x)$ satisfies
$\epsilon(|x|) = 1$ and $\epsilon(-|x|) = -1$.

The matrix element $<B|\bar b(0) \gamma^\beta (1-\gamma_5) U(0,y) b(y)|B>$ which is equal to
$<B|\bar b(0)
\gamma^\beta U(0,y)b(y)|B>$ from parity consideration
contains all information about
initial bound state corrections. It is, however,
difficult to completely evaluate it due to
non-perturbative effects. In the following we attempt two calculations:
one using light cone expansion, and
the other using heavy quark effective theory.

\subsection{Light Cone Expansion Estimates}

In general one can decompose the matrix element,
$<B|\bar b(0)\gamma^\beta U(0,y) b(y)|B>$, in the following form

\begin{eqnarray}
<B|\bar b(0) \gamma^\beta U(0,y)b(y)|B> = 2[P_B^\beta F(y^2, y\cdot P_B) + y^\beta G(y^2,y\cdot P_B)],
\end{eqnarray}
where $F(y^2, y\cdot P_B)$ and $G(y^2,y\cdot P_B)$ are functions of the two
independent Lorentz scalars, $y^2$ and $y\cdot P_B$.

Since we are interested in having large kaon energy $E_K> 2.1$ GeV and small invariant
mass for the $X$,
the dominant contribution to the $y$ integration in Eq.~(\ref{lc})
will be from the light cone region $y^2 \lesssim 1/E^2_K$, which suggestes
that, as a good approximation, $<B|\bar b(0) \gamma^\beta U(0,y) b(y) |B> \approx
2P^\beta_B F(0, y\cdot P_B)$. This approximation is also supported by the fact that
the function $\Delta_{q'}(y)$ has a singularity at $y^2=0$ while away from light cone it
vanishes. Carrying out a Fourier transformation\cite{8},

\begin{eqnarray}
F(0,y\cdot P_B) = \int d\xi e^{-i\xi y\cdot P_B} f(\xi),
\end{eqnarray}
and inserting the above into Eq.~(\ref{lc}), we arrive at

\begin{eqnarray}
&&\sum_X (2\pi)^4\delta^4 (P_B-P_K-P_X) |A(B\to K X)|^2\nonumber\\
&&=
8\pi f^2_K ( 2|\alpha|^2  P^\alpha_K P_K\cdot P_B + |\beta|^2 P_B^\alpha)\nonumber\\
&&\times \int d\xi\,\, \delta [(\xi P_B-P_K)^2] (\xi P_{B\alpha} - P_{K\alpha}) f(\xi).
\end{eqnarray}

We finally obtain the decay  distribution as a function of $E_K$

\begin{eqnarray}
{d\Gamma(B\to K X)\over d E_K}
={f^2_K\over 2\pi m_B} \left (4|\alpha|^2 E_K^2 + |\beta|^2\right ) E_K f({2E_K\over m_B}).
\label{kx}
\end{eqnarray}

Carrying out similar calculations, we obtain the differential decay rate for the $B\to K^* X$
decay

\begin{eqnarray}
{d\Gamma(B\to K^* X)\over d E_{K^*}}
={f^2_{K^*}\over 2\pi m_B} 4|\alpha_*|^2 E_{K^*}^3 f({2E_{K^*}\over m_B}),
\label{kstar}
\end{eqnarray}
where $\alpha_* = (G_F/\sqrt{2})\sum_{q=u,c}V_{qb}V_{qs}^*\tilde A^q$.

It is interesting to note that
the same distribution function $f(\xi)$ appears in both $B\to K X$ and
$B\to K^* X$ cases.
It is also interesting to note that, in the approximation made in this section,
the function $f(\xi)$ is the same as that in
$B\to X \gamma$\cite{8} and semi-leptonic decays
$B\to X l\bar \nu$\cite{10}.
These decays have been studied in details. Experiments
in the future will measure the differential distributions for these decays and, therefore,
provide detailed information about $f(\xi)$. We can use this information in the calculation to
reduce error. One may also turn the argument around to use the decay modes discussed here to
provide constraints on the form of the distribution function $f(\xi)$.
Before the detailed experimental
information becomes available, we
have to make some theoretical modeling for our numerical analysis
which will be discussed later.

\subsection{Heavy Quark Effective Theory Estimates}

We note that the simple expressions for the decay distributions in Eqs.~(\ref{kx})
and (\ref{kstar}) hold to leading order in light cone expansion.
When higher order contributions are included the expressions will
not be so simple.
To have some idea about the sensitivity of the results to other corrections,
in the following we also estimate the corrections to the free $b$ quark decay
rates using heavy quark effective theory.

If the $b$ quark is heavy, the decay products all have large energy and to a good
approximation can be treated as free quarks. In that case,
$U(0,y) \approx  1$ because the
background gluon field can be approximated to vanish,
we then have

\begin{eqnarray}
<B|\bar b(0) \gamma^\beta U(0,y) b(y) |B> \approx <B|\bar b(0) \gamma^\beta b(y)|B>.
\end{eqnarray}
If the $b$ quark is infinitively heavy, the above matrix element is simply
given by
$2P_B^\beta e^{-im_b v\cdot y}$,
where $v$ is the four velocity of the $B$ meson satisfying $v^2 = 1$.
Since the $b$ quark has finite mass, there will be corrections. We now
estimate the leading $1/m_b^2$ corrections following the procedure outlined in
Ref.\cite{9}. In the heavy quark effective theory,
the $b(x)$ quark field can be expanded as

\begin{eqnarray}
b(x) &=& e^{-im_b v\cdot x} \{
1+ i\gamma\cdot D_T/(2m_b) + v\cdot D \gamma\cdot D_T/(4m_b^2)
-(\gamma \cdot D_T)^2/(8m_b^2) \} h(x) + O(1/m_b^3) \nonumber\\
 &+& (\mbox{terms for anti-quark}),\nonumber\\
D^\mu_T &=& D^\mu - v^\mu v\cdot D,\nonumber\\
D^\mu &=& \partial^\mu - i g_s G^\mu(x).
\end{eqnarray}

Using the above expressions and keeping $1/m_b$ terms, we obtain\cite{9}

\begin{eqnarray}
<B|\bar b(0)\gamma^\beta b(y)|B>&=&2m_B
e^{-im_b v\cdot y}
\{ v^\beta - {i\over 6 m_b} (2 y^\beta + v\cdot y v^\beta) (\mu^2_\pi
-\mu^2_g)
\nonumber\\
&-&{1\over 8 } (y^2 - (v\cdot y)^2)v^\beta \mu^2_\pi)\},
\label{heavy}
\end{eqnarray}
where
\begin{eqnarray}
\mu^2_g &=& {1\over 4m_B}
<B|\bar h g_s G_{\mu\nu} \sigma^{\mu\nu} h|B>,
\nonumber\\
\mu^2_\pi &=& - {1\over 2m_B}<B|\bar h (iD_T)^2 h|B>.
\end{eqnarray}
We note that the expansion in Eq.~(\ref{heavy})
is different from the light cone expansion as can be
seen from the above expression that some $y^2$ terms are kept.
The expansion is truncated at order $1/m_b$ in Eq.~(\ref{heavy}). The truncation of the
$1/m_b$ expansion enforces the use of the quark level phase space, instead of the hadron level
phase space.

Inserting the above expression into Eq.~(\ref{rate}), we have

\begin{eqnarray}
\Gamma(B\to K X)
\approx {f^2_K \over 8\pi}m_b [|\alpha|^2m_b^2
(1+{7\over 6}{\mu^2_g\over m_b^2}
-{53\over 6} {\mu^2_\pi\over m_b^2})
+|\beta|^2 (1-{\mu_\pi^2\over 2 m_b^2} + {\mu^2_g \over 2 m_b^2})].
\label{hqet1}
\end{eqnarray}
In the approximation made here, the distribution of $E_K$ is a delta function with
the peak at $E_K = m_b/2$.

Carrying out similar calculations, we obtain the decay rate for the $B\to K^* X$
decay,

\begin{eqnarray}
\Gamma(B\to K^* X)
\approx {f^2_{K^*} \over 8\pi}m_b |\alpha_*|^2m_b^2
(1+{7\over 6}{\mu^2_g\over m_b^2}
-{53\over 6} {\mu^2_\pi\over m_b^2}).
\label{heqet2}
\end{eqnarray}

It is clear that in the limit of large $m_b$, that is $\mu^2_{\pi,g}/m_b^2 \to 0$,
the result reduces to the free
$b$ quark decay $b\to K(K^*) q'$ result as expected.

The expressions for the decay rates, in the approximation we are working with, are
simple, allowing easy analysis.
In the case of the light cone expansion method, one needs to have detailed 
knowdege of 
distribution function $f(\xi)$ for numerical analysis.
Although
the detailed shape is not known, we do know some
properties \cite{8,10}. When integrating $\xi$ from 0 to 1, $\int^1_0d\xi f(\xi)$ must
give 1 due to current conservation.
If the decay can be considered to be a free $b$ quark decay, then $U(0,y) = 1$ because
no background gluon field exists, and the $b$ quark field is given by
$b(y) = e^{-iy\cdot P_b} b(0)$, one obtains

\begin{eqnarray}
f(\xi) = \delta(\xi - {m_b\over m_B}).
\label{new}
\end{eqnarray}

We can also estimate the
mean $<\xi> = \int^1_0 d\xi \xi f(\xi)$ and
the variance $\sigma^2 = \int^1_0 d\xi \xi^2 f(\xi)- \\
<\xi>^2$ using heavy quark effective theory. They are given by\cite{10,8}

\begin{eqnarray}
&&<\xi> = {m_b\over m_B} [1+ {5\over 6m^2_b} (\mu^2_\pi-\mu^2_g)],\nonumber\\
&&\sigma^2 = {\mu^2_\pi\over 3m_B^2}.
\end{eqnarray}
The small value for $\sigma^2$ implies that the distribution function is sharply peaked around
$m_b/m_B$.

To go further we take the following parameterization for the distribution
function\cite{10}

\begin{eqnarray}
f(\xi) = N {\xi (1-\xi)^c\over [(\xi -a)^2 +b^2]^d},
\label{dis}
\end{eqnarray}
where $N$ is a normalization constant which guarantees
$\int^1_0 d \xi f(\xi) = 1$. This function reduces to a $\delta$-function
with the peak at $a$ as $b\to 0$. Comparing with Eq. (\ref{new}), in this limit
$a=m_b/m_B$.
Once the parameters $c$ and $d$ are given, the parameters $a$ and $b$ can be
fixed by comparing with $<\xi>$ and $\sigma^2$. Unfortunately we do not know the values for
$c$ and $d$ at present. We will take $c$ and $d$
to be free parameters and vary them to see how the energy spectra of $K(K^*)$,
branching ratios and
CP asymmetries are changed.

\section{Results and discussions}

We are now ready to present our numerical analysis.
We will make theoretical predictions for the kaon energy spectra
$d\Gamma/d E_{K^{(*)}}$, CP-averaged
branching ratios and direct CP asymmetries defined as

\begin{eqnarray}
Br_{ave}(B\to K(K^*) X)&=& {1\over 2}[Br(B \to K(K^*) X) + Br( \bar B \to \bar K(\bar K^*) \bar X)],\nonumber\\
A_{CP}(B\to K(K^*) X) &=& {\Gamma(B\to K(K^*) X) - \Gamma(\bar B\to \bar K (\bar K^*) \bar X)
\over \Gamma(B\to K(K^*) X) + \Gamma(\bar B\to \bar K(\bar K^*) \bar X)}.
\end{eqnarray}

For the numerical analysis,
we need to know the values for the parameters involved.
Some of them are well determined.
In our numerical calculations we will use the following values for the
relevant parameters \cite{PDG}:
$m_b = 4.9$ GeV, $m_c = 1.5$ GeV, $m_s = 120$ MeV, $m_d = 4$ MeV,
$m_u= 2$ MeV,
$|V_{us}| = 0.2196$, $|V_{cb}| = 0.0402$, $|V_{ub}/V_{cb}| = 0.085$,
$f_K = 160$ MeV, $f_{K^*} = 214$ MeV, $\alpha_s(M_Z) = 0.118$.
We keep the CP violating phase $\gamma$ to be a free parameter and vary it to see how the
branching ratios  and CP asymmetries depend on it.

The HQET parameter $\mu^2_g$ can be extracted from the $B^\ast-B$ mass
splitting: $\mu^2_g = 3(m^2_{B^\ast}-m^2_B)/4 \simeq 0.36$ GeV$^2$,
while $\mu^2_\pi$ is less determined. A calculation of QCD sum rules
gives $\mu^2_\pi = (0.5 \pm 0.2)$ GeV$^2$ \cite{sumr}, which is consistent with
$\mu^2_\pi = (0.45 \pm 0.12)$ GeV$^2$ from a recent lattice QCD
calculation \cite{latt}.
We will use $\mu^2_\pi = 0.5$ GeV$^2$ for our numerical calculations.

In the case of light cone expansion, we also need to specify the
distribution function $f(\xi)$. We will assume it to be the
form given in Eq.~(\ref{dis}).
To have some idea how the kaon energy spectra, branching ratios, and CP
asymmetries depend on the form of the
distribution function, we consider two very different forms \cite{form}:
(i) preset $c=d=1$, in that case $a=0.9548$ and $b=0.005444$ determined by
the known mean value and variance of the distribution function;
(ii) preset $c=d=2$, in that case $a = 0.9864$ and $b=0.02557$ determined by
the same mean value and variance of the distribution function.

In Figs.~1-4, we show the kaon energy spectra in $B\to K(K^\ast) X$ decays
computed in the light cone expansion approach, assuming $\gamma=60^\circ$.
The solid and dashed curves correspond, respectively, to the parameter
set (i) and (ii) for the distribution function.
The kaon energy spectra are a discrete line at $E_{K^{(\ast)}}=m_b/2$ in free
$b$ quark decay approximation, which is not shown in the figures.
We see that initial bound state effects stretch the spectra over the full
kinematic range $0\leq E_{K^{(\ast)}}\leq m_B/2$ and the kaon energy spectra
depend strongly on the form of the distribution function.
However, we note that all the spectra have more than $97\%$ of
events with $E_{K^{(\ast)}}>2.1$ GeV. This implies that if
the integrated branching ratios and CP asymmetries are measured
with $E_{K^{(\ast)}}>2.1$ GeV,
the effects from the detailed shape of the distribution function are small.

We show the CP-averaged branching ratios, in Figs.~5-8, and the CP asymmetries, in Figs.~9-12,
in $B\to K(K^\ast) X$ as a function of the CP violating phase
$\gamma$. The solid curves are the results from the light cone
expansion using the parameter set (i) for the distribution function, while
the dashed curves are from the free $b$ quark decay approximation.
The initial bound state effects
encoded in the distribution function almost cancel completely
in the CP asymmetries in
$B\to K^\ast X$,
so that the solid and dashed curves coincide in Figs.~11 and 12.
We find that the shifts in the branching
ratios and CP asymmetries are negligible
if the parameter set (ii) instead of (i) for the distribution function is
used, indicating that both the branching ratios and
the CP asymmetries are insensitive to
the detailed shape of the distribution function.

One can clearly see from Figs.~5-8 that the differences between the solid and dashed
curves are
small, about 2\%. This implies that according to light cone expansion
estimates, the initial bound state effects increase the CP-averaged
branching ratios for $B\to K(K^\ast) X$ by about $2\%$, largely because
the $B\to K(K^\ast) X$ phase space is used, which is larger than the $b\to K(K^\ast) q^\prime$
phase space used in the free $b$-quark and heavy quark effective theory calculations.
The branching ratios for $\bar B^0 \to  K^-(K^{\ast -}) X$ are sensitive to
$\gamma$, varying from $0.53(0.25) \times 10^{-4}$ to $1.5(2.0)\times 10^{-4}$,
whereas the
branching ratios for $B^-\to \bar K^0(\bar K^{\ast 0}) X$ are not sensitive to
$\gamma$,
varying from $0.77(0.67)\times 10^{-4}$ to $0.84(0.74)\times 10^{-4}$.
The above sensitivities
to $\gamma$ can be easily understood by noticing that the tree operators
$O_{1,2}$ contribute to $\bar B^0\to K^-(K^{\ast -}) X$ decays
but not to $B^-\to \bar K^0(\bar K^{\ast 0}) X$ decays when small annihilation
contributions are neglected, resulting
in strong dependence on $V_{ub}V_{us}^*$ for the former, but not for the
latter.

For the same reasons, the CP asymmetries are expected to be
much larger in $\bar B^0\to K^-(K^{\ast -}) X$ than in
$B^- \to \bar K^0(\bar K^{\ast 0}) X$. The differences between
the solid curves and dashed curves 
in Figs.~9 and 10 are very small, about $1\%$.
This implies that according to light cone expansion estimates, the
initial bound state effects increase the CP asymmetries in
$B\to KX$ by about $1\%$.
They do not affect the CP asymmetries in $B\to K^* X$.
The CP asymmetries in $\bar B^0 \to K^-(K^{\ast -}) X$
can be as large as $7\%(14\%)$, but very small ($< 1\%$) in
$B^- \to \bar K^0(\bar K^{\ast 0}) X$, as expected.

The heavy quark effective theory estimates of the initial bound
state effects are always to reduce the branching ratios at the level of
10\% as can be seen from Eqs.~(\ref{hqet1}) and (\ref{heqet2}) if
$\mu^2_\pi = \mu^2_g$ is used.
In fact within the allowed range for $\mu^2_\pi$ the initial bound
state effects tend to reduce the branching ratios.
The CP asymmetries are the
same as those obtained by free b quark decay approximation.

The three estimates (free quark decay approximation, light cone expansion and
heavy quark effective theory method)
carried out here all give the same order of magnitudes for the branching
ratios and CP asymmetries which are also the same order of
magnitudes as those obtained in Ref.\cite{2}. The initial bound state effects are
at the order of 10\% of the free b quark decay estimates. The differences 
between different methods may be viewed as uncertainties in the estimates.
The branching ratios are of order $10^{-4}$ and are within the reach of the
$B$ factories. The CP asymmetries in the neutral $B$ modes $\bar B^0 \to K^-(K^{*-}) X$
are large and can be measured at the $B$ factories. When
more data become available, one may obtain interesting information about
hadronic effects and also information about the CP violating phase $\gamma$.

\acknowledgments
This work was supported in part by NSC
under grant number NSC 89-2112-M-002-058 and by the Australian Research
Council.

\begin{figure}[htb]
\centerline{ \DESepsf(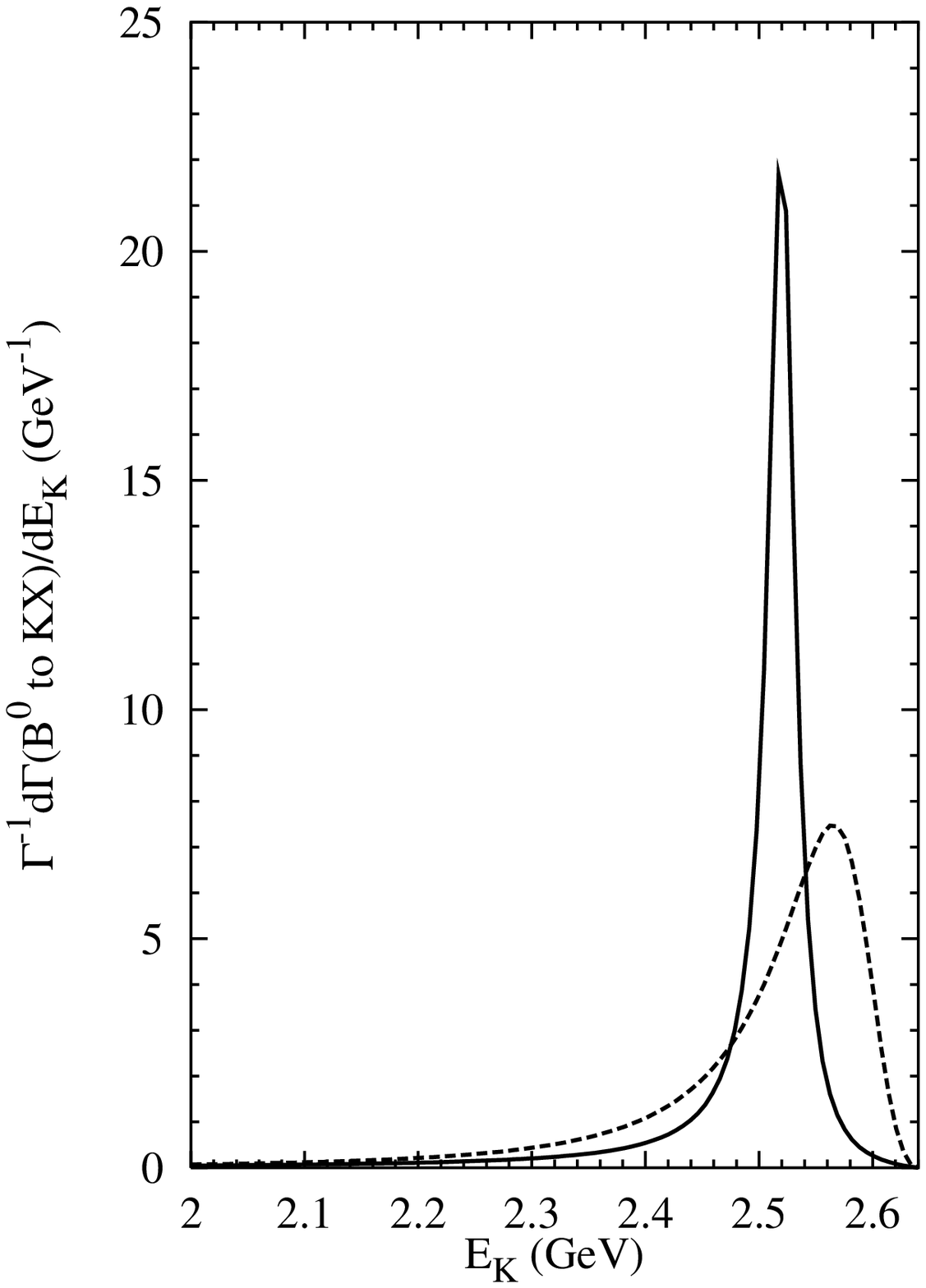 width 10cm)}
\smallskip
\caption {
Kaon energy spectrum in $\bar B^0 \to K^- X$.
In Figs.~1-4, the solid curves are for (i) $c=d=1$; the
dashed curves are for (ii) $c=d=2$.} \label{b0sp}
\end{figure}

\begin{figure}[htb]
\centerline{ \DESepsf(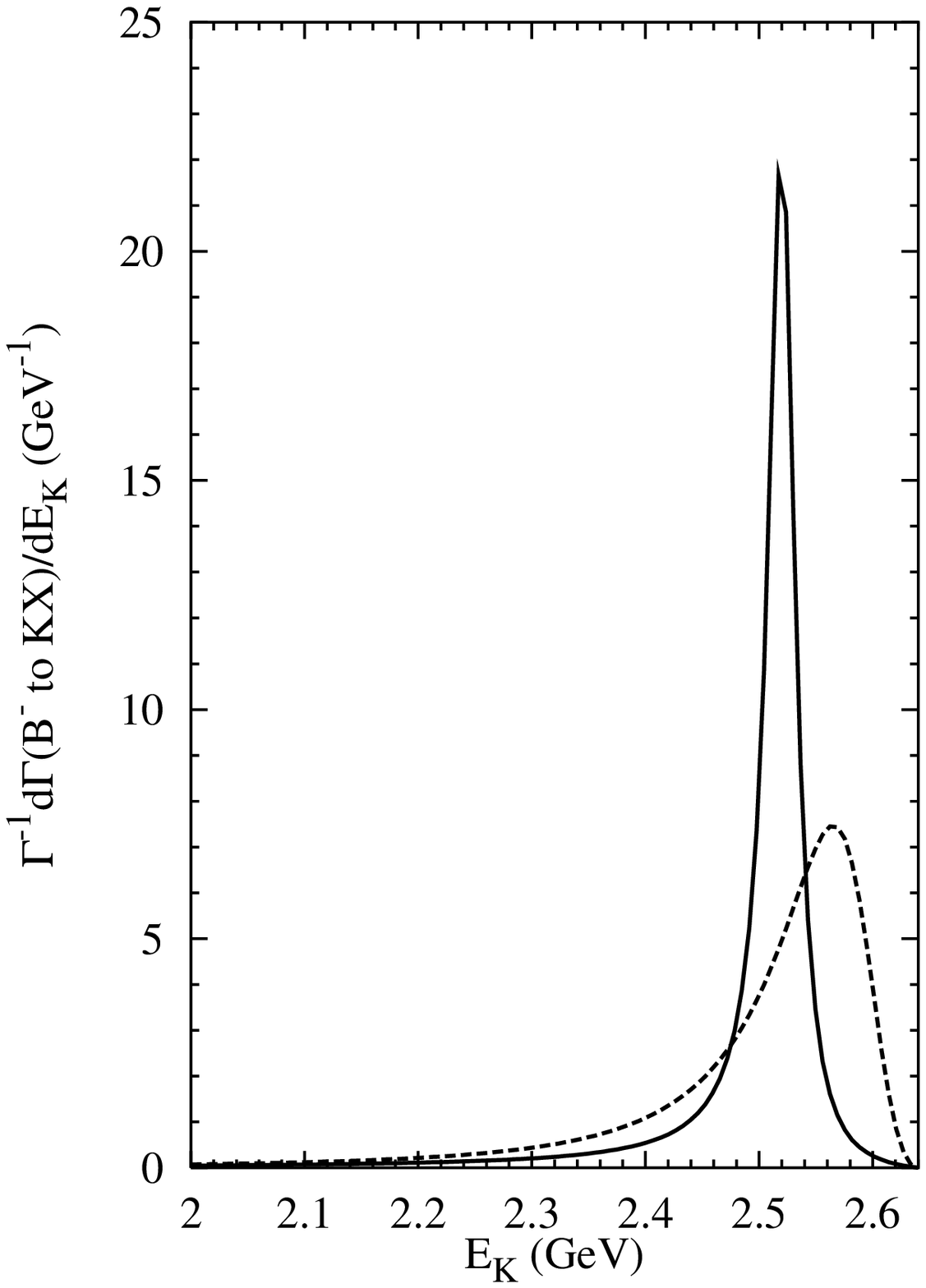 width 10cm)}
\smallskip
\caption {Kaon energy spectrum in $B^-\to \bar K^0 X$.
} \label{Bmsp}
\end{figure}

\begin{figure}[htb]
\centerline{ \DESepsf(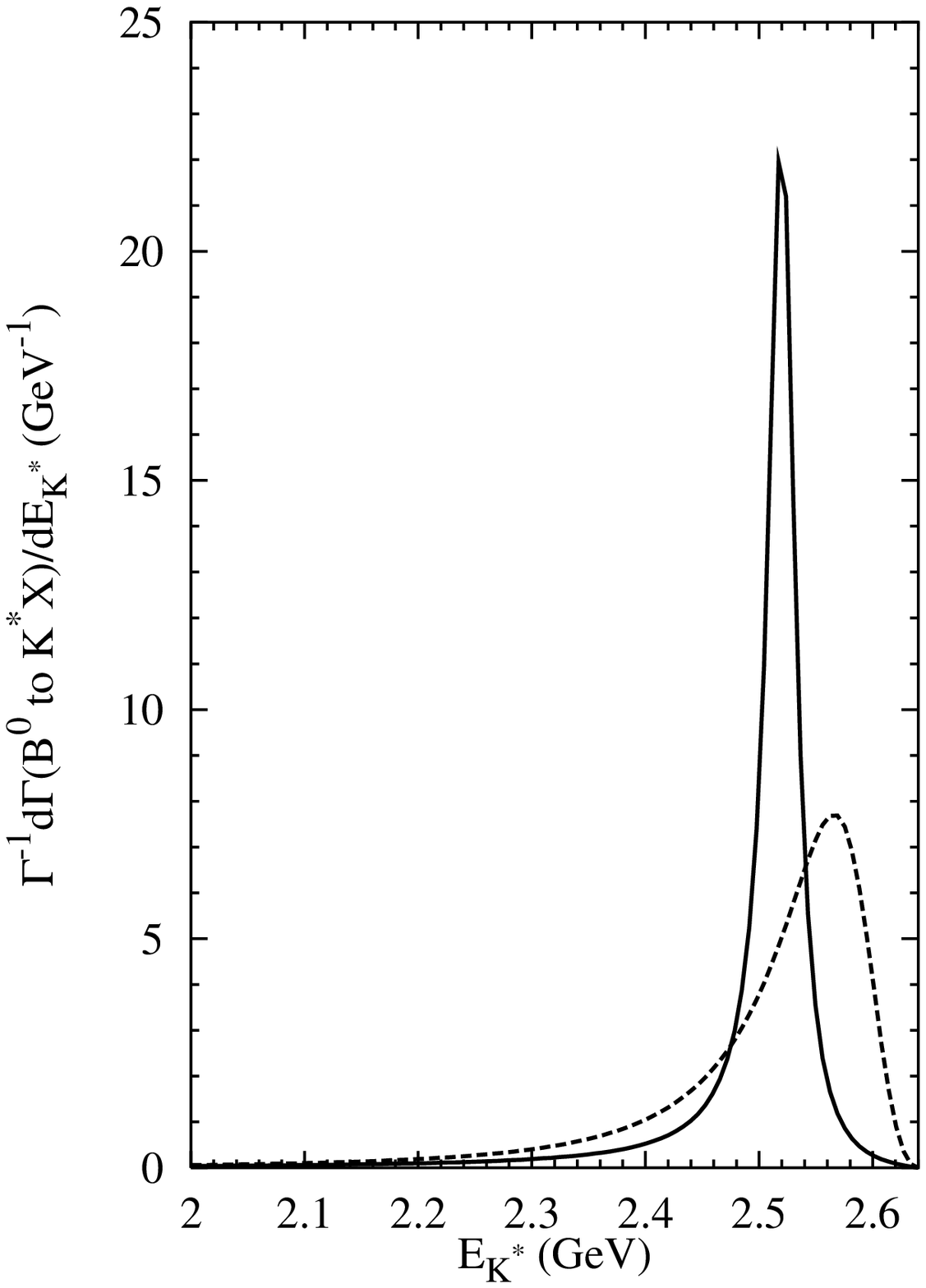 width 10cm)}
\smallskip
\caption {
Kaon energy spectrum in $\bar B^0 \to K^{\ast -} X$.
} \label{b0starsp}
\end{figure}

\begin{figure}[htb]
\centerline{ \DESepsf(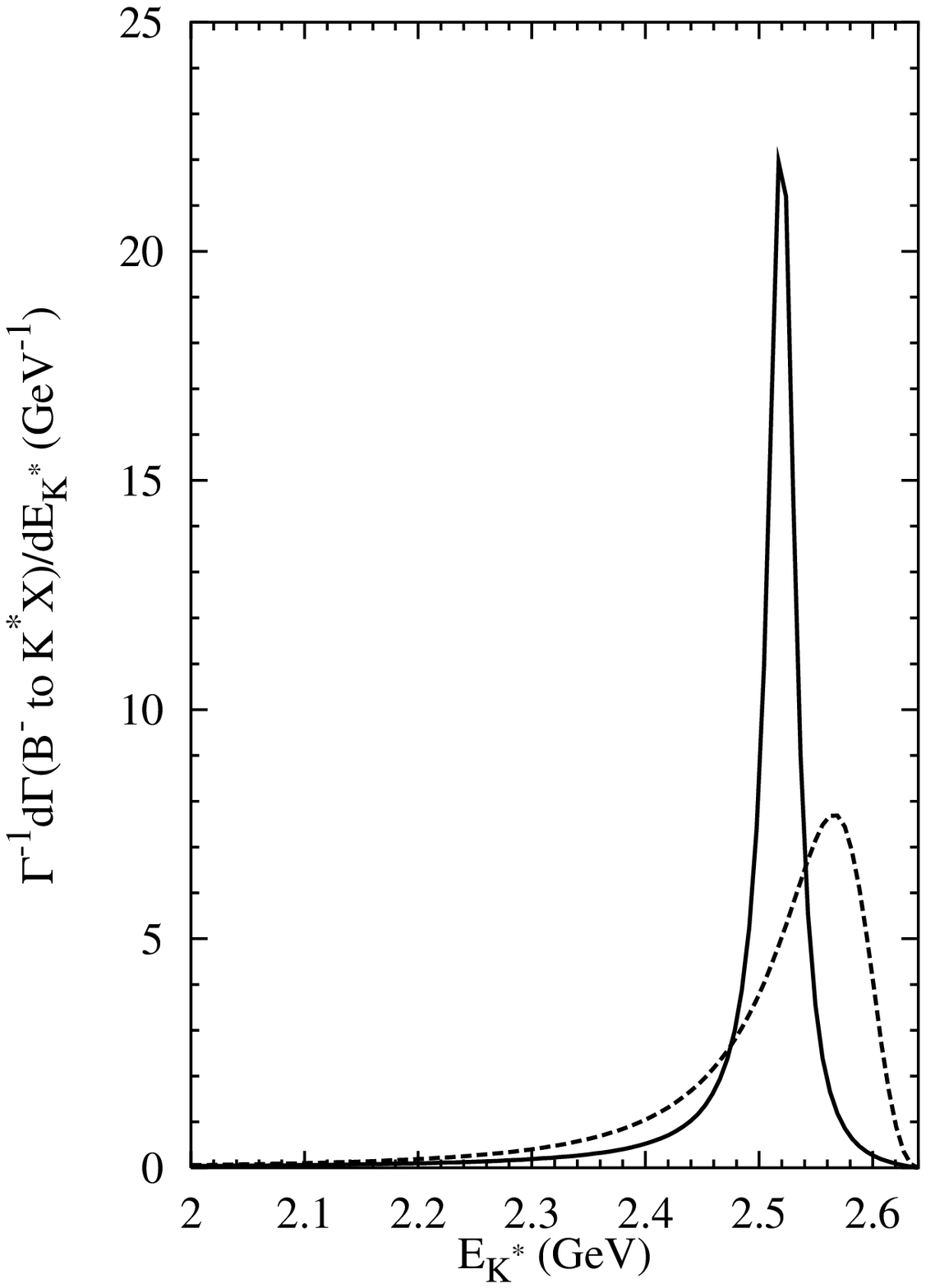 width 10cm)}
\smallskip
\caption {Kaon energy spectrum in $B^-\to \bar K^{\ast 0} X$.
} \label{Bmstarsp}
\end{figure}

\newpage
\begin{figure}[htb]
\centerline{ \DESepsf(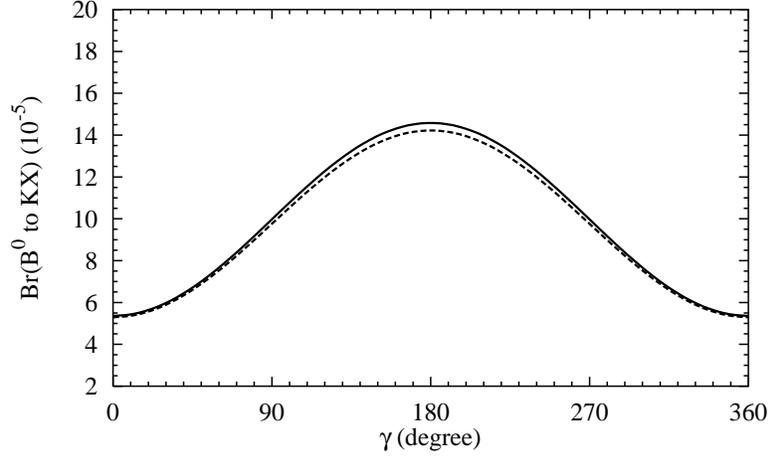 width 10cm)}
\smallskip
\caption {
CP-averaged branching ratio for $\bar B^0 \to K^- X$.
In Figs.~5-10, the solid curves are for the ligh cone expansion with
(i) $c=d=1$; the dashed curves are for the free $b$ quark decay approximation.}
\label{b0br}
\end{figure}

\begin{figure}[htb]
\centerline{ \DESepsf(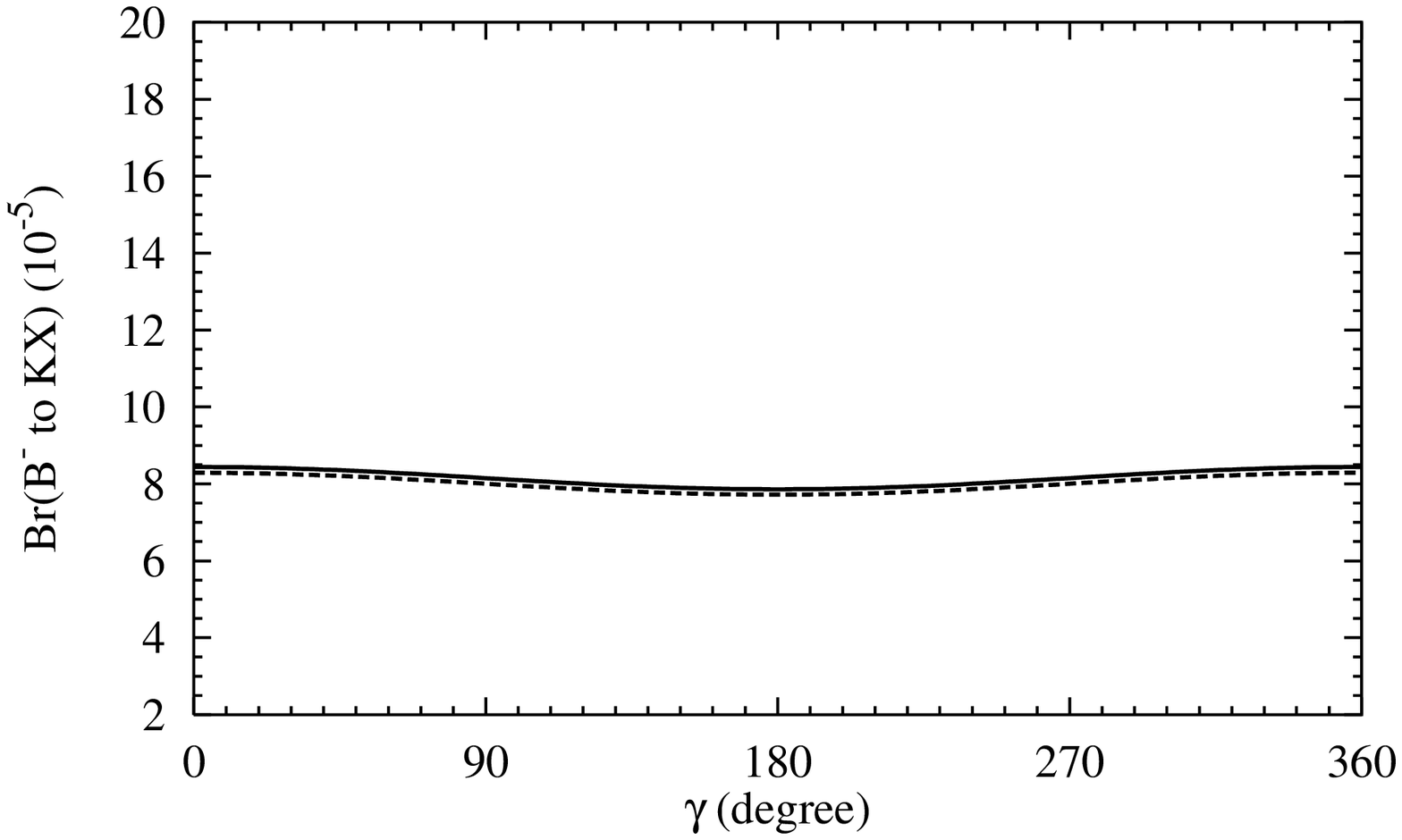 width 10cm)}
\smallskip
\caption {
CP-averaged branching ratio for $B^-\to \bar K^0 X$.
} \label{bmbr}
\end{figure}

\begin{figure}[htb]
\centerline{ \DESepsf(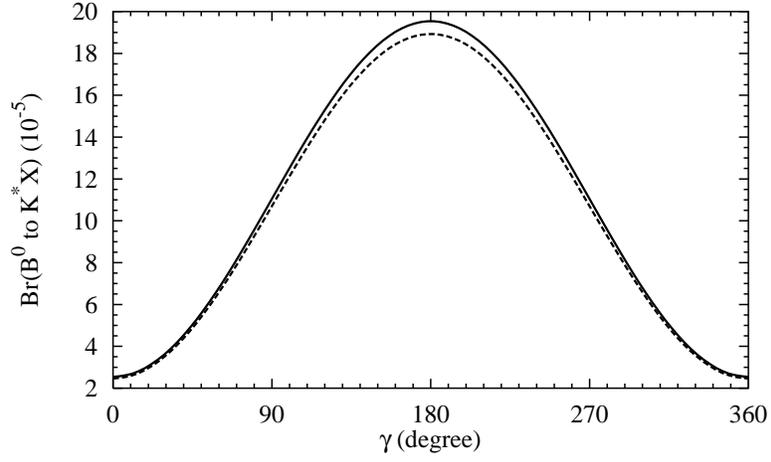 width 10cm)}
\smallskip
\caption {
CP-averaged branching ratio for $\bar B^0 \to K^{*-} X$.
} \label{b0starbr}
\end{figure}

\begin{figure}[htb]
\centerline{ \DESepsf(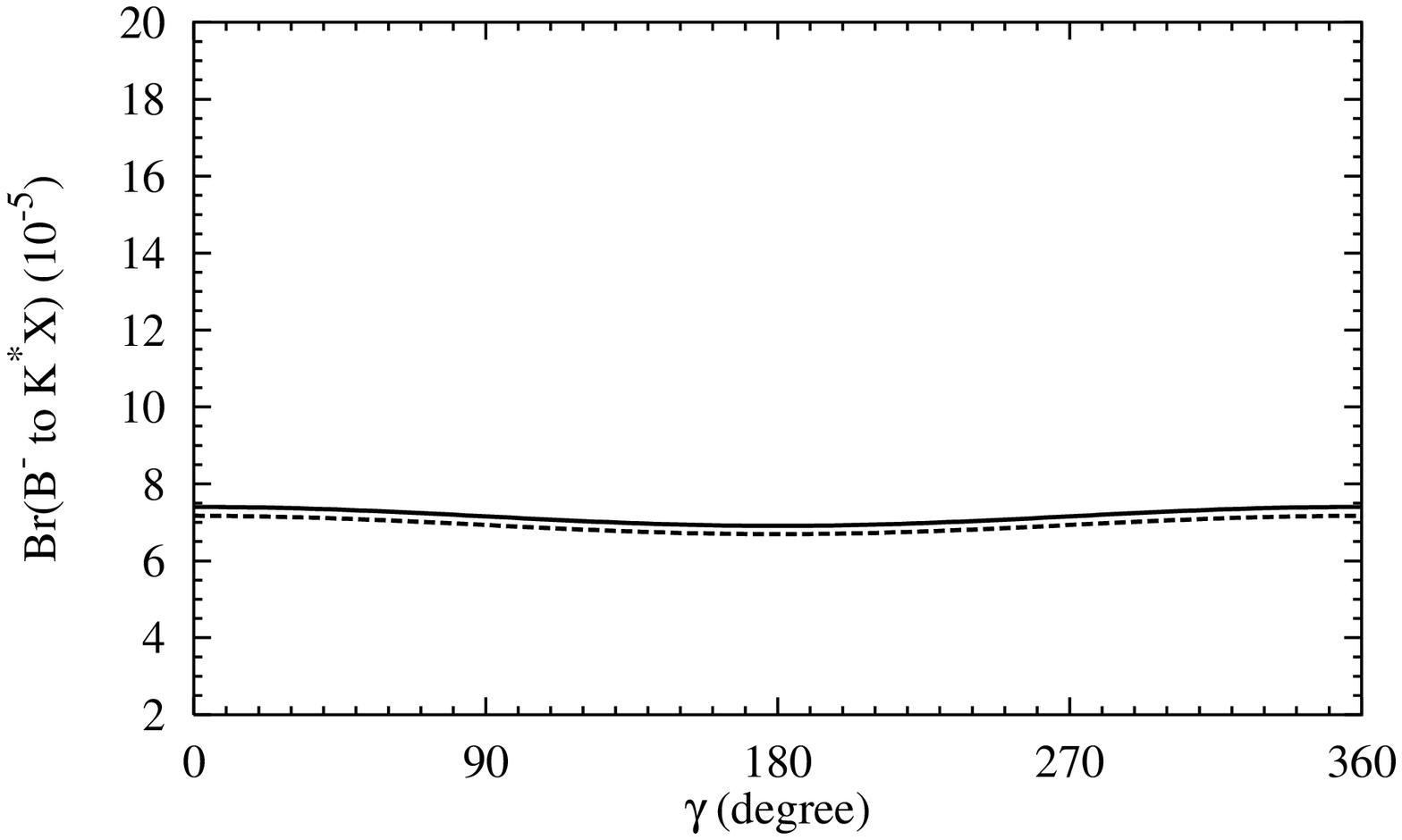 width 10cm)}
\smallskip
\caption {
CP-averaged branching ratio for $B^-\to \bar K^{*0} X$.
} \label{bmstarbr}
\end{figure}

\newpage
\begin{figure}[htb]
\centerline{ \DESepsf(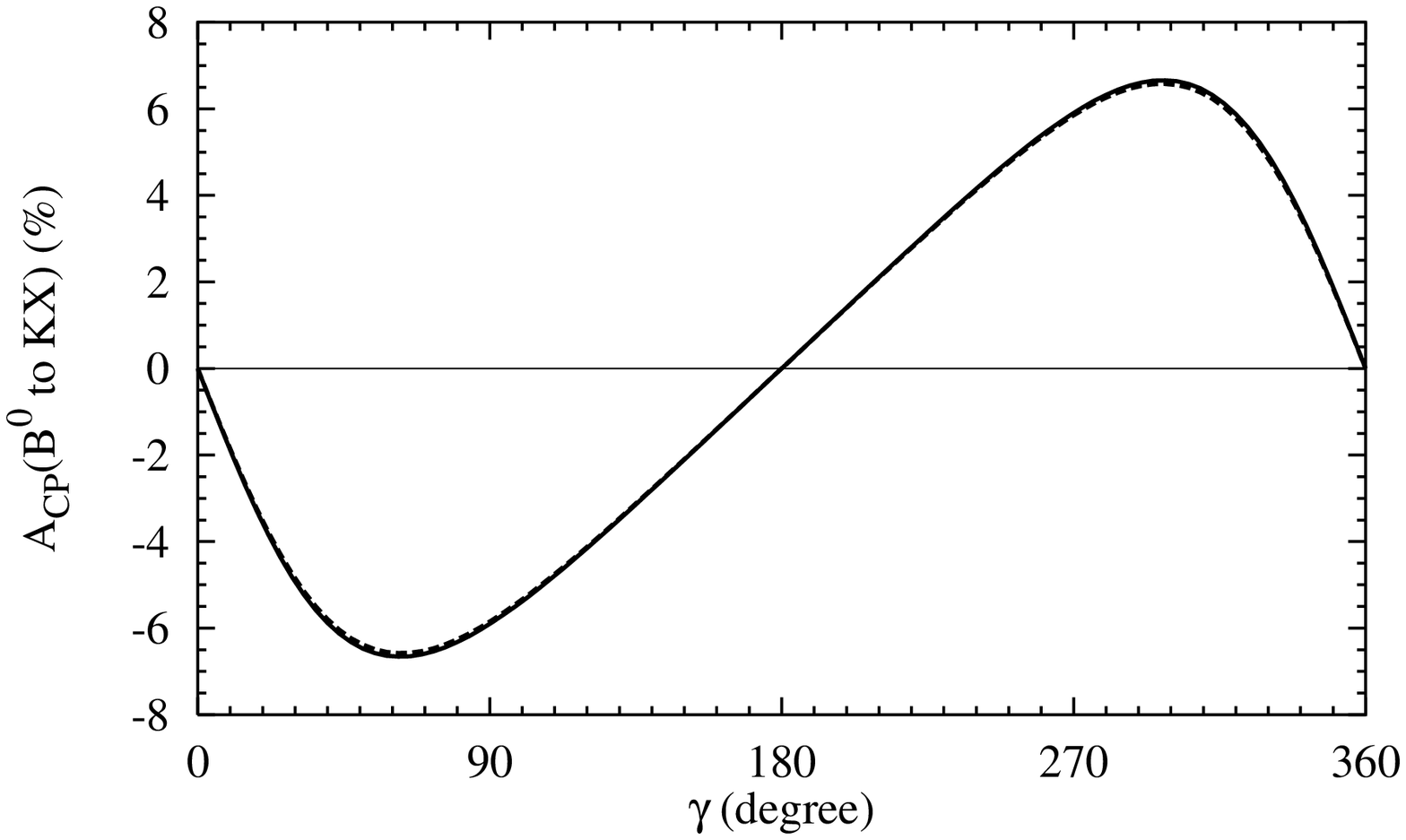 width 10cm)}
\smallskip
\caption {
CP asymmetry in $\bar B^0 \to K^- X$.
} \label{b0cp}
\end{figure}

\begin{figure}[htb]
\centerline{ \DESepsf(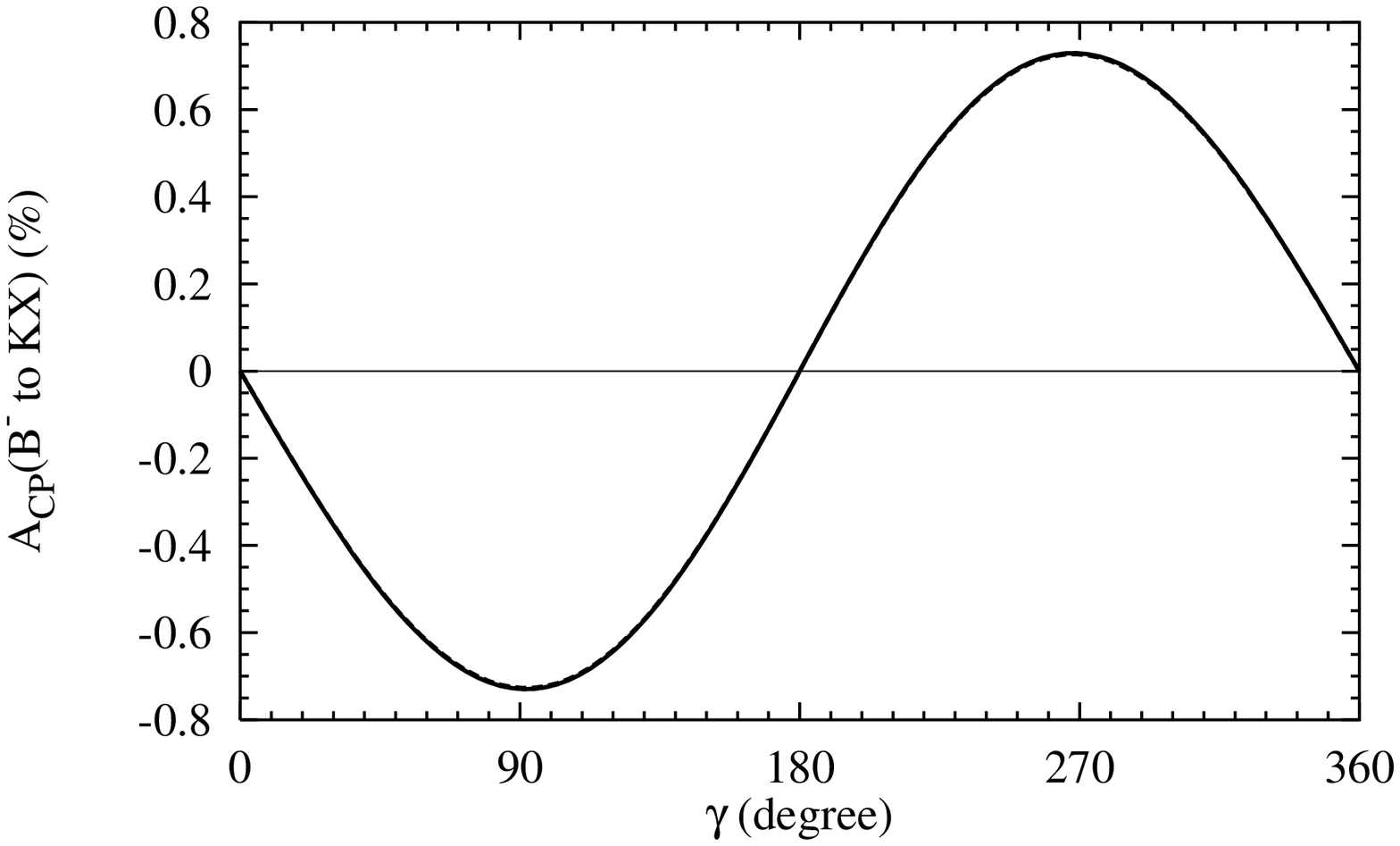 width 10cm)}
\smallskip
\caption {
CP asymmetry in $B^-\to \bar K^0 X$.
} \label{bmcp}
\end{figure}

\newpage
\begin{figure}[htb]
\centerline{ \DESepsf(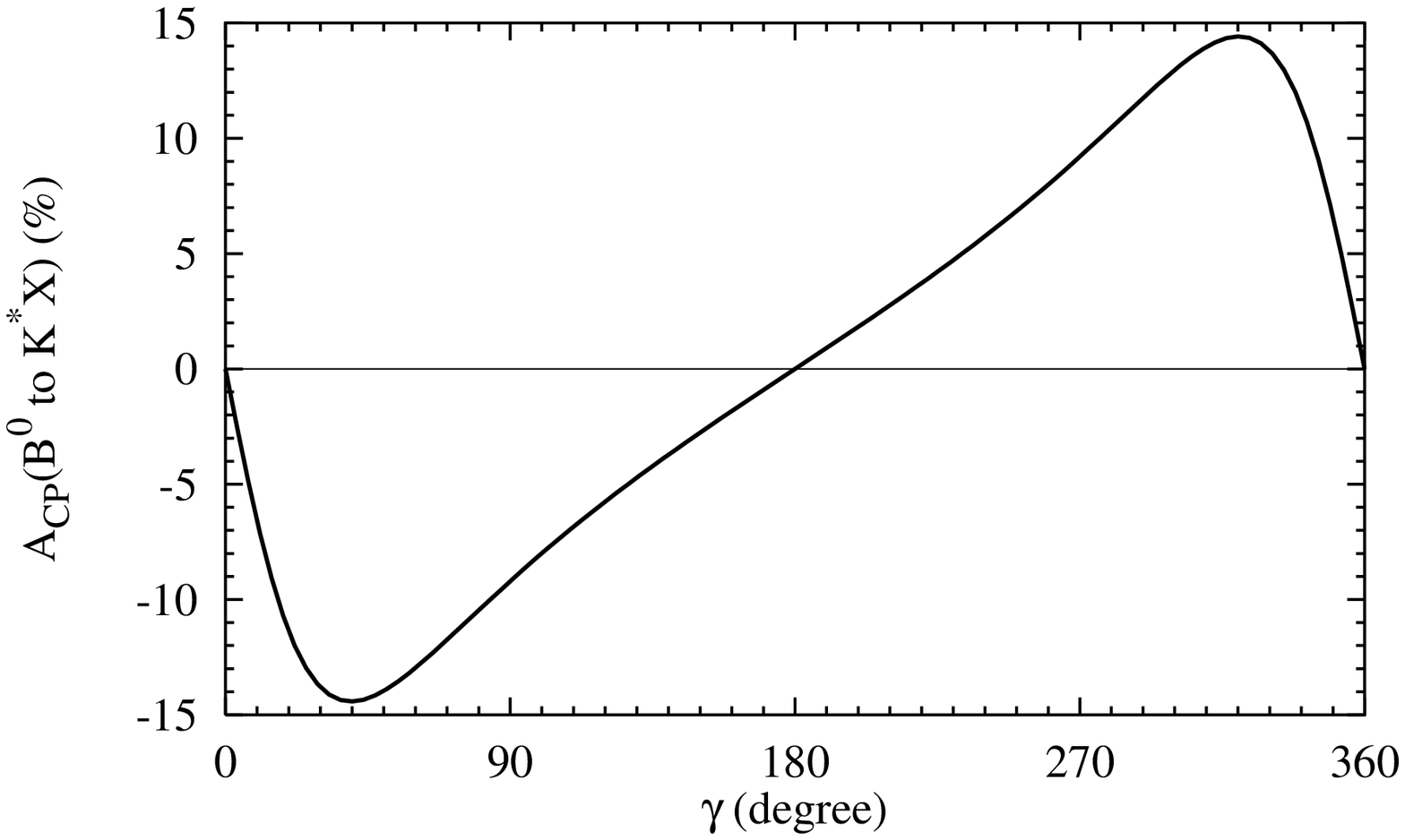 width 10cm)}
\smallskip
\caption {
CP asymmetry in $\bar B^0 \to K^{*-} X$.
} \label{b0starcp}
\end{figure}

\begin{figure}[htb]
\centerline{ \DESepsf(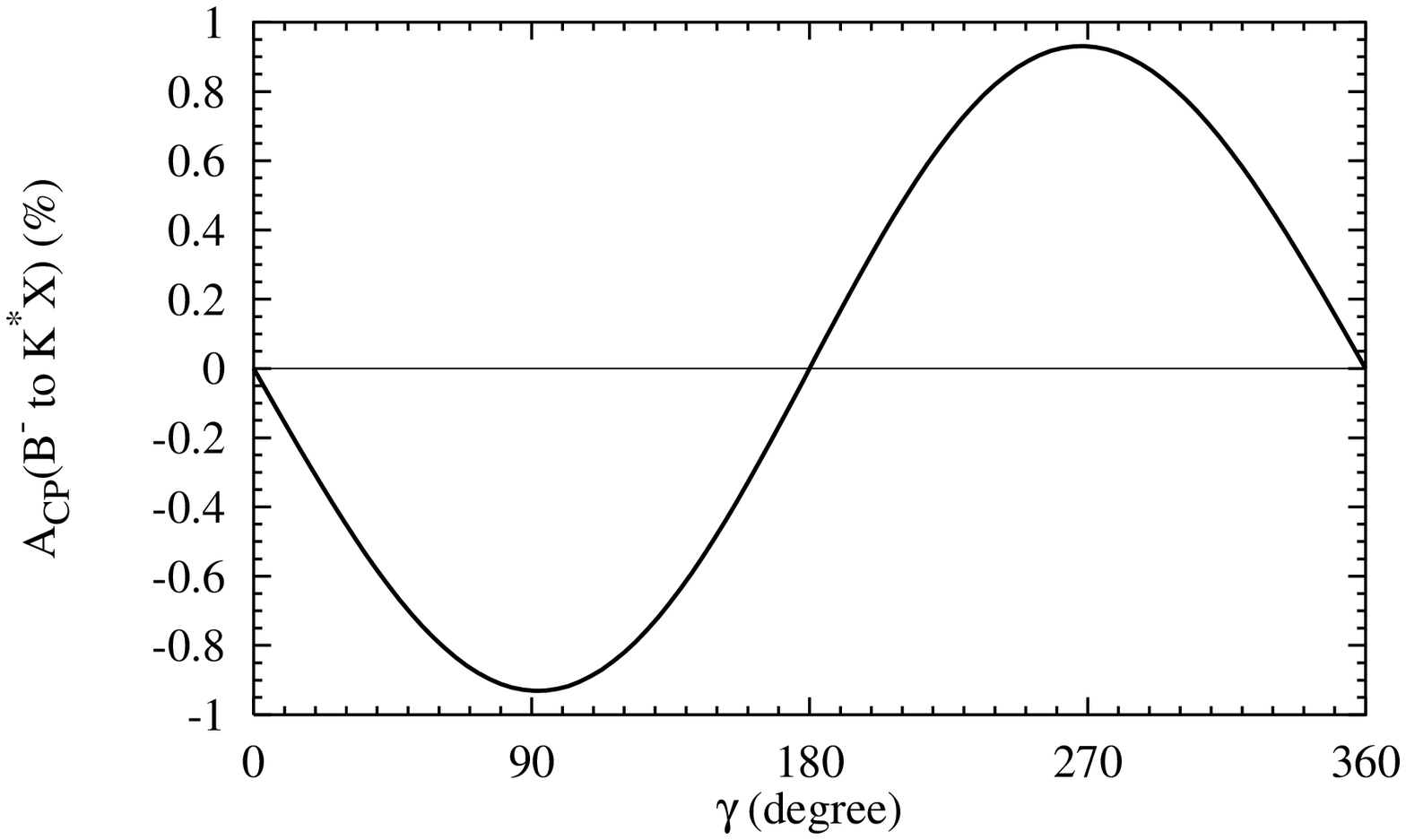 width 10cm)}
\smallskip
\caption {
CP asymmetry in $B^-\to \bar K^{*0} X$.
} \label{bmstarcp}
\end{figure}


\begin{references}
\bibitem{1}
H.-Y. Cheng and K.-C. Yang, e-print hep-ph/9910291;
G. Deshpande et al., Phys. Rev. Lett. {\bf 82}, 2240 (1999);
X.-G. He, W.-S. Hou and K.-C. Yang, Phys. Rev. Lett. {\bf 81}, 5738 (1998).
A. Ali, G. Kramer and C.-D. Lu, Phys. Rev. {\bf D58 }, 094009 (1998);
G. Deshpande, X.-G. He and J. Trameptic, Phys. Lett. {\bf B 345}, 547 (1995).

\bibitem{2}
A. Datta et al., Phys. Rev. {\bf D 57}, 6829 (1998).

\bibitem{3}
X.-G. He, J.P. Ma and C.-Y. Wu, e-print hep-ph/0008159;
A. Datta, X.-G. He and S. Pakvasa, Phys. Lett. {\bf B 419}, 369 (1998);
D. Atwood and A. Soni, Phys. Rev. Lett. {\bf 81}, 3324 (1998);
A. Kagan and A. Petrov, e-print hep-ph/9707354.

\bibitem{4} A. Szczepaniak, E. Henley and S. Brodsky,
Phys. Lett. {\bf B 243}, 287 (1990);
H.-n. Li and H.-L. Yu, Phys. Rev. Lett. {\bf 74}, 4388 (1995);
M. Beneke et al., Phys. Rev. Lett. {\bf 83}, 1914 (1999);
Nucl. Phys. {\bf B 591}, 313 (2000).

\bibitem{5}
Y.-Y. Keum, H.-n. Li and A.I. Sanda, e-print hep-ph/0004004, 0004173;
C.-D. Lu, K. Ukai and M.-Z. Yang, e-print hep-ph/0004213;
D.-S. Du, D.-S. Yang and G.-H. Zhu, Phys. Lett. {\bf B 488}, 46 (2000);
e-print hep-ph/0008216;
T. Muta, A. Sugamoto, M.-Z. Yang and Y.-D. Yang,
Phys. Rev. {\bf D 62}, 094020 (2000);
M.-Z. Yang and Y.-D. Yang, Phys. Rev. {\bf D 62}, 114019 (2000).

\bibitem{6} G. Buchalla, A. Buras and M. Lautenbacher,
Rev. Mod. Phys. {\bf 68}, 1125 (1996);
A. Buras, M. Jamin and M. Lautenbacher, Nucl. Phys. {\bf B 400}, 75 (1993);
M. Ciuchini et al., Nucl. Phys. {\bf B 415}, 403 (1994);
N. Deshpande and X.-G. He, Phys. Lett. {\bf B 336}, 471 (1994).

\bibitem{7}
N. G. Deshpande et al., Phys. Lett. {\bf B 366}, 300 (1996).

\bibitem{8} C.H. Jin, Eur. Phys. J. {\bf C 11}, 335 (1999).

\bibitem{9} J.P. Ma, Phys. Lett. {\bf B 488}, 55 (2000).

\bibitem{10} C.H. Jin and E.A. Paschos, in {\it Proceedings of the
International
Symposium on Heavy Flavor and Electroweak Theory}, Beijing, China,
1995, edited by C.H. Chang and C.S. Huang (World Scientific, Singapore,
1996), p.~132; e-print hep-ph/9504375;
C.H. Jin, Phys. Rev. {\bf D56}, 2928 (1997);
C.H. Jin and E.A. Paschos, Eur. Phys. J. {\bf C 1}, 523 (1998);
C.H. Jin, Phys. Rev. {\bf D 56}, 7267 (1997);
Phys. Rev. {\bf D 62}, 014020 (2000).

\bibitem{PDG} Particle Data Group, D.E. Groom et al.,
Eur. Phys. J. {\bf C 15}, 1 (2000).

\bibitem{sumr} P. Ball and V.M. Braun, Phys. Rev. {\bf D 49}, 2472 (1994).

\bibitem{latt} A.S. Kronfeld and J.N. Simone,
Phys. Lett. {\bf B 490}, 228 (2000).

\bibitem{form} C.H. Jin, Phys. Rev. {\bf D 57}, 6851 (1998).


\end{references}
\end{document}